# Automated Allocation of Detention Rooms Based on Inverse Graph Partitioning

Jingwei Wang*, Chuan Liu*, Yukai Zhao, Yunlong Ma, Min Liu, and Weiming Shen, *Fellow, IEEE*

*Abstract*— Room allocation is a challenging task in detention centers since lots of related detainees need to be held separately with limited rooms. It is extremely difficult and risky to allocate rooms manually, especially for organized crime groups with close connections. To tackle this problem, we develop an intelligent room allocation system for detention centers to provide optimized room allocation schemes automatically. We first formalize the detention room allocation problem as inverse graph partitioning, which can measure the quality of room allocation schemes. Then, we propose two heuristic algorithms to achieve the global optimization and local optimization of detention room allocation. Experiment results on real-world datasets show that the proposed algorithms significantly outperform manual allocation and suggest that the system is of great practical application value.

## I. Introduction

Room allocation is a crucial task in detention centers as detainees with close relations need to be held separately to reduce risks. According to reports from detention centers, if detainees with social relationships are assigned to the same rooms, they could collaborate with each other and be highly likely to oppose the management of detention centers, such as bullying others, cooperating in concealing the crime, and even cooperating in a crime after their release. Therefore, keeping the related detainees separately facilitates the management of detention centers and reduces the risk of recidivism. However, it is very difficult to divide all related detainees into different rooms due to the limited room resources of detention centers, especially for organized crime groups with close connections [1].

At present, detention center managers allocate rooms to detainees manually relying on their experience, which makes this problem even more salient. First, it is difficult for detention center managers to sort out all social relations per person due to the large number of detainees. As a result, some related detainees who could have been separated are held together and form groups with close connections, leading to potential risks mentioned before. Second, it takes detention center managers two or three days per week to adjust rooms of detainees, which is laborious and inefficient. More importantly, manual allocation cannot optimize the rooms of detainees from global perspective, because it is impossible to find out how many related detainees are held in the same room.

Essentially, detention room allocation (DRA) aims to minimize the number of relations of detainees inside rooms of detention centers. This is a limited resource allocation problem. One common approach is to turn it into a classic combinatorial optimization problem over graphs. Motivated by the approaches described in [2], [3], we formalize the DRA problem as inverse graph partitioning which is the inverse problem of graph partitioning. Then, we propose two heuristic algorithms to achieve the global optimization and local optimization of room allocation. Based on that, we develop an intelligent room allocation system for detention centers to provide optimized room allocation schemes automatically. An overview of this proposed system is shown in Fig. 1.

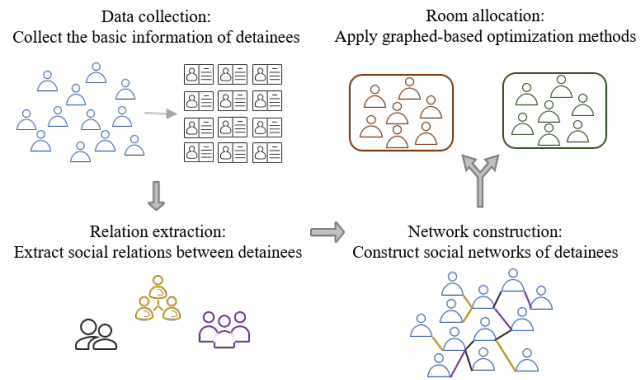

Figure 1. Overview of the propsoed system.

The main contributions of this work are summarized as follows:

- We formalize the DRA problem as inverse graph partitioning and build a mathematical model for it. The objective function of inverse graph partitioning can evaluate the quality of room allocation schemes.

- We design an intelligent room allocation system that can collect data of detainees, extract relations between detainees, construct social networks, and provide room allocation schemes automatically.

*Jingwei Wang and Chuan Liu contributed equally to this work.

Jingwei Wang, Yukai Zhao, Yunlong Ma, and Min Liu are with College of Electronics and Information Engineering, Tongji University, Shanghai 201804, China (e-mail: jwwang@tongji.edu.cn; zhaoyukaijake@tongji.edu.cn; evanma@tongji.edu.cn; lmin@tongji.edu.cn). Corresponding author: Yunlong Ma (phone: +86 137-611-61573; e-mail: evanma@tongji.edu.cn).

Chuan Liu and Weiming Shen are with State Key Laboratory of Digital Manufacturing Equipment and Technology, Huazhong University of Science and Technology, Wuhan 430074, China (e-mail: chuan_liu@ hust.edu.cn; wshen@ieee.org).

- We propose two heuristic algorithms to address the DRA problem, including *hub first assignment* algorithm (HFA) and *local greedy assignment* algorithm (LGA).
- We conduct experiments on real-world datasets to show the effectiveness of the proposed algorithms in comparison with manual allocation.

## II. RELATED WORK

Combinatorial optimization problems over graphs arising from numerous application domains, such as social networks [4], transportation [5], telecommunications [6], and scheduling [7], have thus attracted considerable attention over the years [8]. Resource allocation is an important application of combinatorial optimization. One of the common paradigms is to turn a resource allocation problem into a classic graph theory problem, and solve it with graph-based approaches. Jiang and Hao [9] used the graph coloring method to allocate frequency points in Vehicular Ad-hoc Network. Wang and Zhang [10] implemented the joint resource allocation by solving the max-flow problem on satellite networks. With the Internet of Things (IoT) technology, Kumar and Zaveri [11] utilized a maximum bipartite graph approach to allocate emergency resources during disaster situations. To solve the channel allocation problem in mobile communication, Yao and Yin [12] applied graph coloring methods from the microscopic perspective.

It is worth noting that graph partitioning is related to our problem, which has recently gained great attention due to its application in clustering and community detection [13], [14]. Graph partitioning was first proposed by Kernighan and Lin in 1970 [15], which arises in several physical situations, such as assigning electronic circuits' components to circuit boards to reduce connections between boards [16]. After that, it has been widely used to parallel computing, task scheduling, traffic forecasting, to name a few [17]–[21]. Graph partitioning is a combinatorial optimization problem and falls under the category of NP-hard problems. A strictly exhaustive procedure for finding the optimal partition is often out of the question. Solutions are generally derived using heuristics and approximation algorithms. The first heuristic algorithm is the Kernighan-Lin algorithm which is specified for balanced graph partitioning that is partitioning a graph into two subgraphs of equal size [15]. Li and Zhang [22] investigated the unbalanced graph partitioning problem where subgraphs have unequal sizes and proposed an approximation algorithm to solve it. With the increasing scale of graphs, Li et al. [23] recently adopted a genetic algorithm for large-scale graph partitioning. Besides, many streaming graph partitioning methods have been developed in recent years [24], [25].

In this paper, we model the DRA problem as inverse graph partitioning which is the inverse problem of graph partitioning. To the best of our knowledge, our work is the first to explore inverse graph partitioning. Similar to graph partitioning, inverse graph partitioning is also a combinatorial optimization problem. Most combinatorial optimization problems over graphs are NP-hard, and exact algorithms are time-consuming. Heuristic algorithms have shown promising results in this field, which are regarded as effective solutions. Therefore, we consider using heuristic algorithms to solve the DRA problem.

## III. INTELLIGENT ROOM ALLOCATION SYSTEM

Here we propose a graph-based intelligent room allocation system for detention centers to provide an optimized room allocation scheme automatically. As shown in Fig. 1, the system consists of following four parts: data collection, relation extraction, network construction, and room allocation. Next, we introduce the four parts in turn.

The first step is data collection that is to obtain the basic information of detainees in detention centers, including their rooms, genders, case numbers, crime types, birth places, and main social relations. The information is desensitized and the personal attributes are coded by discrete numbers, so as to protect the privacy of detainees.

The next step is to extract the relations of detainees from the collected data. In addition to the previous social relations between the detainees, three kinds of relations are concerned by detention centers, namely, the relationship of joint crime, the relationship of fellow-townsman, and the relationship of the same type of crime. Detainees having the relationship of joint crime are accomplices who committed a crime together. The relationship of fellow-townsman means that two detainees come from the same town or city. The relationship of the same type of crime refers to those who are detained for the same reason although they may not have committed a crime together. These relationships can be automatically extracted from the collected data.

After obtaining the relationships between detainees, we can use them to construct social networks of the detainees in a detention center. Specifically, if two detainees have any of the above relationships, there is a link between them in the network. Here, we do not consider the weight and the type of links in the network. Since detainees are usually separated by sex in each detention center, there are two networks for one detention center. We use capital letters to denote the detention center and use M and F to denote males and females, respectively. Then, the name of a network is named by the letter of the detention center and the gender of detainees. For example, AM and AF represent the male network and the female network of detention center A, respectively. Here, we visualize AF network, as shown in Fig. 2. The size of a node is proportional to its degree which is the number of links connected to it. The color is used to distinguish different rooms.

Based on the social network of detainees, we can measure the quality of different room allocation schemes using the number of relations between detainees inside rooms. Hence, the goal of the DRA problem turns to reduce the number of relations between detainees inside rooms as much as possible. Here, we formalize the DRA problem as inverse graph partitioning, propose two heuristic algorithms to address this problem, and realize automated allocation of detention rooms. This part will be described in detail in the next section.

Note that all the above steps can be realized automatically through programming without manual participation. Therefore, the system can provide an optimized room allocation scheme for detention centers automatically.

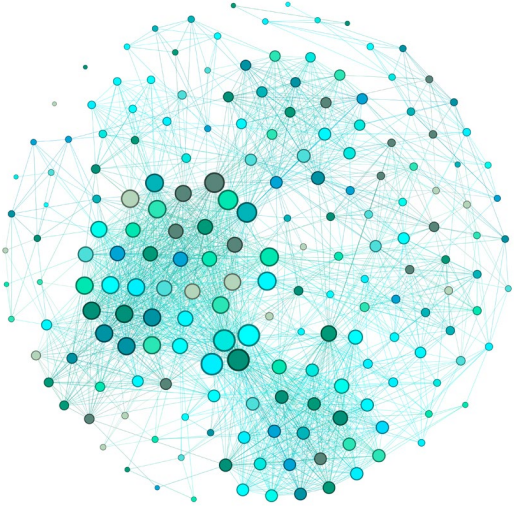

Figure 2. AF network.

## IV. OPTIMIZATION MODEL AND ALGORITHMS

In this section, we formalize the DRA problem as inverse graph partitioning and build a mathematical model for it. Then, two heuristic algorithms are proposed to implement global optimization and local optimization of detention room allocation, respectively.

### A. Graph-based Optimization Model

Detention room allocation aims to divide all the detainees into a predefined number of rooms and minimize the connections of detainees inside rooms. From the perspective of graph theory, this problem is equivalent to the inverse problem of graph partitioning which divides a graph into a predefined number of smaller subgraphs and minimizes the number of edges between subgraphs. Therefore, we turn the detention room allocation problem into an inverse graph partitioning problem which is to divide a graph into many subgraphs and minimizes the number of edges within subgraphs. In this way, the optimal solution to inverse graph partitioning is the best room allocation scheme for detention centers.

We first formalize the inverse graph partitioning problem. Given an undirected network $G(V, E)$ in which $V$ represents the set of nodes (detainees) and $E$ represents the set of links (relations). The number of nodes and links is denoted by $|V|$ and $|E|$, respectively. Let $K$ represent the number of subgraphs (rooms). The size of subgraph $k$ is denoted by $s_k$. The number of nodes and links of subgraph $k$ are denoted by $|V_k|$ and $|E_k|$, respectively, where $k = 1,…, K$. All links in $G$ can be categorized into two classes: one is the links inside subgraphs (intra-room links), $E_{in}$, the other is the links outside subgraphs (inter-room links), $E_{out}$. The goal of inverse graph partitioning is to minimize the number of intra-room links $|E_{in}|=\Sigma|E_k|$. Then, the objective function is

$$f = \Sigma |E_k|. \qquad (1)$$

In fact, detention room allocation has to satisfy two hard constraints: (1) each detainee (node) must and can only be assigned to a room that is $V_1 \cup … \cup V_K = V$ and $V_i \cup V_j = \Phi$, $\forall i \neq j$. (2) the number of detainees inside each room cannot exceed its capacity that is $|V_i| \leq s_i$, $i = 1,…, K$. In this paper, we assume all rooms having the same capacity that is $s_1=…= s_K = S := \lceil |V|/K \rceil$, where $\lceil |V|/K \rceil$ is the smallest integer not less than $|V|/K$. Note that the number of rooms must be limited and less than the number of nodes; otherwise, the best scheme is that each room contains only one node and $|E_{in}|$ reaches the minimum (i.e., 0). Clearly, the objective function $f$ can quantify the quality of room allocation schemes.

Inverse graph partitioning is a combinatorial optimization problem on graphs, which is an NP-hard problem. A strictly exhaustive procedure for finding the best solution is often out of the question. Therefore, we devise two heuristic algorithms that can provide the near-optimal solution and have lower time complexity than exact algorithms. Among them, hub first assignment algorithm (HFA) serves as the global optimization of detention room allocation, which aims to optimize the room allocation of all detainees. Local greedy assignment algorithm (LGA) is used to adjust some detainees' rooms, which is the local optimization and more common in practice.

### B. Global Optimization

*Hub first assignment* algorithm (HFA) is motivated by the solution of the graph coloring problem [26]. It is well-known that a common feature of real-world networks is the presence of hubs, which means a few nodes are densely connected to other nodes in the network. Those nodes with a high degree are called hubs and those with a low degree are named peripheral nodes. The degree distribution will have a long tail since the presence of hubs, indicating hubs have a much higher degree than most other nodes. Therefore, a natural idea is to give priority to hubs, that is to allocate the room to those detainees with a large number of connections at first.

Specifically, all nodes are listed in descending order according to their degrees at the starting point. Then, the head node of the list is selected out and moved to any room $k$. If the node has connections with other nodes in room $k$, one computes the number of those links, denoted by $\Delta|E_k| = |\Gamma_k(v)|$, where $\Gamma_k(v)$ presents the set of neighbors of node $v$ in room $k$. If the number of nodes exceeds the size of room $k$, we assign a very large value to $\Delta|E_k|$. An array of $K$ increments $\Delta|E_1|,…, \Delta|E_k|$ is obtained when the node is moved to the $K$ rooms in turn. Then this node is assigned to the room with the minimum $\Delta|E|$. If there are two or more rooms corresponding to the minimum $\Delta|E|$, we randomly choose one from them to move the head node into it. After that, deleting the head node from the list reduces the number of nodes from $n$ to $n-1$ and then a new round of assignment could be started until all nodes are removed. The procedure ensures that each assignment is the best but it may not lead to the optimal solution in the end. It is highly likely that a near-optimal solution is found by HFA algorithm, which is much better than manual allocation. The time complexity of HFA is $O(n^2)$.

**Algorithm 1.** *Hub first assignment* algorithm
**Input:** Graph $G(V, E)$, $K$ rooms with size $S$
**Output:** Node assignment
1: Sort $V$ in descending order of node degree
2: **for** $v \in V$ **do**
3:    **for** $k = 1,…, K$ **do**
3:      **if** $|V_k| > S$ **then**
4:        $\Delta|E_k| \leftarrow |E|$
5:      **else**
6:        $\Delta|E_k| \leftarrow |\Gamma_k(v)|$
7:      **end if**
8:    **end for**
9:    Select the minimum $\Delta|E_k|$
10:   Assign node $v$ to room $k$
11: **return** node assignment

### C. Local Optimization

The global optimization usually provides a near-optimal room allocation scheme for detention centers. However, a large number of detainees need to be rearranged to new rooms according to the solution given by HFA, leading to huge labor costs, which is unfeasible in a real situation. Hence, we propose a local optimization algorithm that only a small number of detainees are rearranged each time to reduce the labor cost. In this way, the goal turns to maximize the optimization effect with the limited number of adjustments.

To realize this goal, a *local greedy assignment* algorithm (LGA) is developed in this work. Assuming that only $m$ detainees can be adjusted at a time, the algorithm will choose the one that makes the biggest drop each time. Specifically, we first find out $m$ detainees who have the most intra-room links and push them into a virtual queue. The algorithm traverses the virtual queue and calculates the maximum reduction of intra-room links that can be achieved by moving a detainee to another room. Each time we adjust the room of the individual who can lead to the greatest reduction in intra-room links. Then, a new individual with the most intra-room links among the rest of the detainees is added into the virtual queue, and a new round of selection begins until the number of adjustments is reached. In practice, the value of $m$ can be set by users. This flexible allocation method meets the real-life needs of detention centers better. The time complexity of LGA is $O(mn)$.

**Algorithm 2.** *Local greedy assignment* algorithm
**Input:** Graph $G(V, E)$, adjusted number $m$
**Output:** Node assignment
1: $Q \leftarrow$ pick $m$ nodes with the most $E_{in}$ from $V$
2: **for** $i = 1,…, m$ **do**
3:    **for** $v$ in $Q$ **do**
3:      **for** $k = 1,…, K$ **do**
4:        $\Delta|E_k| \leftarrow$ move $v$ to room $k$
5:      **end for**
6:      Select the minimum $\Delta|E_k|$ of node $v$
7:    **end for**
8:    Select the node $v$ with minimum $\Delta|E_k|$ among $Q$
9:    Rearrange node $v$ to subgraph $k$
10:   Remove node $v$ from $Q$
11:   Add a new node with the most $E_{in}$ from $V$ to $Q$
12: **return**

### V. RESULTS

We use four networks obtained from two real detention centers for experiments. The basic topological features of the four networks are summarized in Table 1. <k> represents the average degree of nodes and CC represents the number of connected components in a network. Note that the detention network maybe unconnected because there are few links between crime groups and there may be some isolated nodes. Generally, females are fewer than males in detention centers so that the number of rooms of females is only about half of that of males. Here, we assume that all rooms of a detention center have the same capacity that is the maximum number of detainees a room can hold. When a room is full, detainees

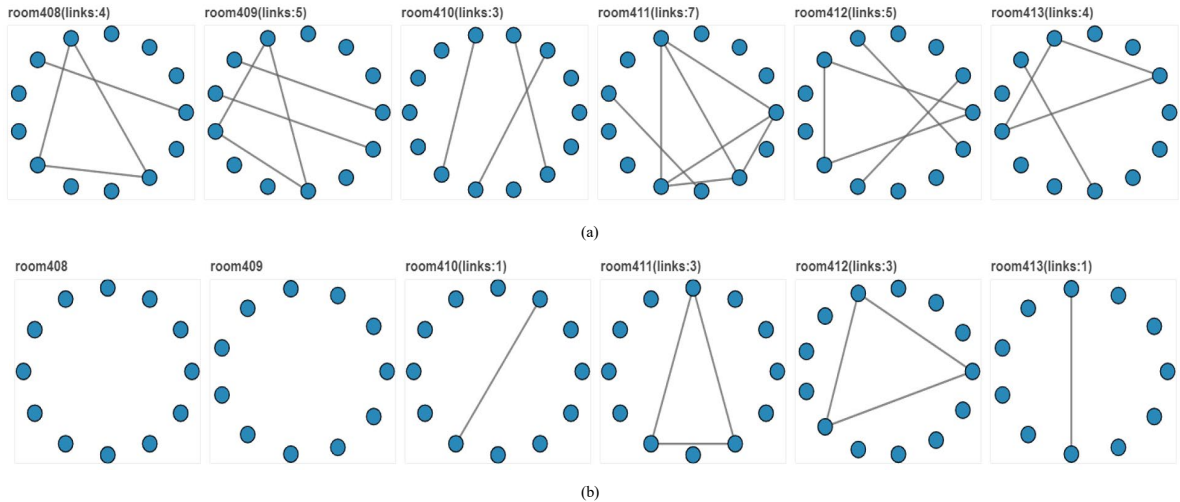

Figure 3. The real intra-room links between people in six rooms of AF network before and after optimization. (a) represents the real intra-room links before optimization, (b) represents the real intra-room links after optimization.

from other rooms can no longer be assigned to the room.

TABLE I. BASIC TOPOLOGICAL FEATURES OF FOUR DETENTION NETWORKS

| Datasets | Basic Topological Features | | | | |
|---|---|---|---|---|---|
| | Nodes | Links | <k> | CC | Rooms |
| AM | 903 | 48719 | 107.904 | 2 | 44 |
| AF | 190 | 2119 | 22.305 | 3 | 16 |
| BM | 680 | 25697 | 75.579 | 10 | 34 |
| BF | 273 | 4182 | 30.637 | 16 | 17 |

We compare our graph-based allocation methods with traditional manual allocation. Here, we use the number of intra-room links as the evaluation metric.

Table 2 shows the results of global optimization on four networks. On average, global optimization achieves an improvement of 50% over manual allocation. The huge reduction confirms that manual allocation cannot separate detainees who could have been separated. Fig. 3 shows the real intra-room links between detainees in six rooms of AF network before and after optimization, which are provided by our intelligent room allocation system. We can observe a significant decrease in the number of intra-room links. This visualization clearly shows the effectiveness of the proposed system.

TABLE II. NUMBER OF INTRA-ROOM LINKS IN FOUR NETWORKS

| Methods | Datasets | | | |
|---|---|---|---|---|
| | AM | AF | BM | BF |
| Manual | 1177 | 124 | 914 | 323 |
| Global Optimization | 608 | 51 | 448 | 135 |

Figs. 4-7 show the results of local optimization on four networks. The number of adjustments varies among different networks. It is obvious that the number of intra-room links decreases with the increase of the number of adjustments and the downward trend is near linear in all networks. When the number of adjustments reaches a certain number, the results of local optimization approach to global optimization results and cannot be improved further. It also indicates that there is no need to adjust the room for all detainees. By rearranging rooms for about a quarter of the detainees, we can achieve similar results as global optimization.

The results of global optimization and local optimization show that the proposed room allocation algorithms have great advantages over manual allocation. Therefore, the proposed system can provide optimized room allocation schemes automatically and efficiently, saving manpower and time.

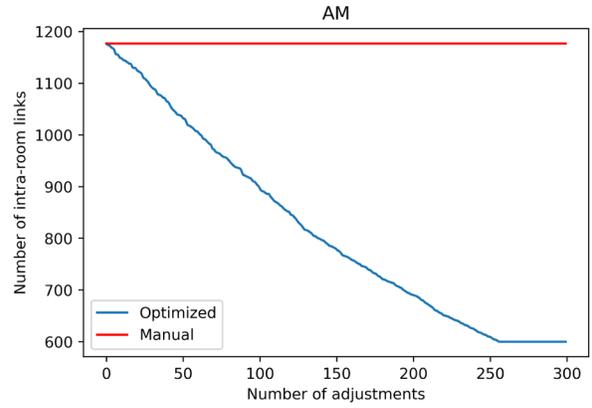

Figure 4. The result of local optimization allocation in AM network.

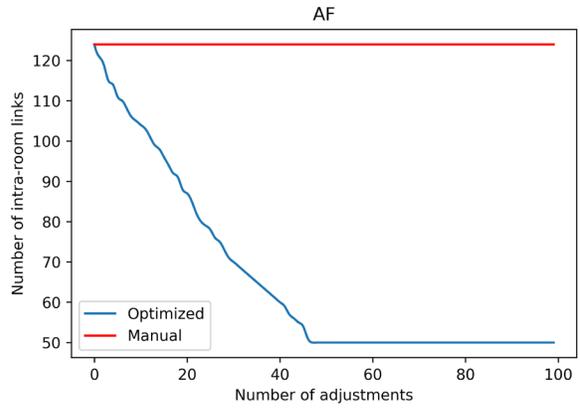

Figure 5. The result of local optimization allocation in AF network.

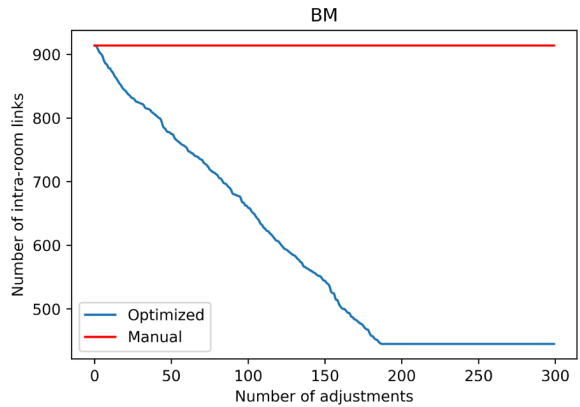

Figure 6. The result of local optimization allocation in BM network.

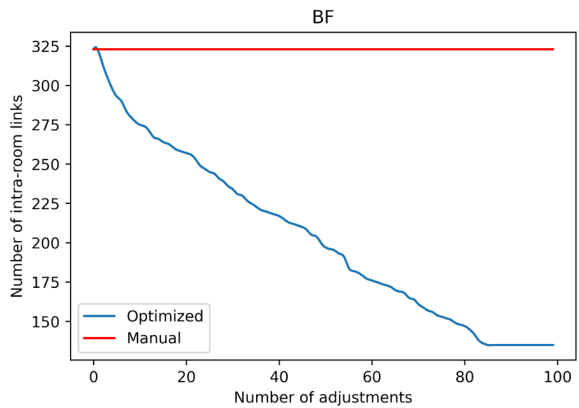

Figure 7. The result of local optimization allocation in BF network.

## VI. Conclusion

In this paper, we present an intelligent room allocation system for detention centers to efficiently and reasonably assign detainees to limited rooms with the goal of making intra-room links the least. The primary contribution of this work is that we model the detention room allocation problem as a combinatorial optimization problem over graphs that is the inverse graph partitioning, which could measure the effect of different room allocation schemes. Beyond that, we propose two heuristic algorithms to allocate rooms for detainees automatically, thus obviating the need for manual adjustments which are laborious and unpromising. Results on two real detention centers demonstrate the effectiveness of the proposed algorithms and suggest that the system is of great practical application value.

## Acknowledgment

This work was supported in part by the National Key R&D Program of China under grant 2019YFB1704700, the National Natural Science Foundation of China under grants 61573257, 71690234, 61873191, 61973237, and 72074170, and the Science and Technology Commission of Shanghai Municipality under grants 19JG0500700 and 20JG0500200.